\documentclass[prl,aps,twocolumn,superscriptaddress]{revtex4-2}

\usepackage{bm}
\usepackage{amsmath}
\usepackage[T1]{fontenc}
\usepackage{xspace}
\usepackage{comment}
\usepackage{bbold}
\usepackage{braket}
\usepackage{ascmac}

\ifx\pdfoutput\undefined
\usepackage[dvipdfmx]{graphicx}
\usepackage[dvipdfmx]{hyperref}
\usepackage[dvipdfmx]{color}
\usepackage[dvipdfmx]{xcolor}
\else
\usepackage{graphicx}
\usepackage{hyperref}
\usepackage{color}
\usepackage{xcolor}
\fi

\graphicspath{
{./}
}

\begin{document}

\title{Crystallization dynamics of magnetic skyrmions in a frustrated itinerant magnet}

\author{Kotaro Shimizu}
\affiliation{Department of Applied Physics, The University of Tokyo, Tokyo 113-8656, Japan} 
\affiliation{Department of Physics, University of Virginia, Charlottesville, Virginia 22904, USA}

\author{Gia-Wei Chern} 
\affiliation{Department of Physics, University of Virginia, Charlottesville, Virginia 22904, USA}

\date{\today}

\begin{abstract}
We investigate the phase ordering kinetics of skyrmion lattice (SkL) in a metallic magnet. The SkL can be viewed as a superposition of magnetic stripes whose periods are determined by the quasi-nesting wave vectors of the underlying Fermi surface. An effective magnetic Hamiltonian that describes the electron-mediated spin-spin interaction is obtained for a two-dimensional s-d model with the Rashba spin-orbit coupling.  Large-scale Landau-Lifshitz-Gilbert dynamics simulations based on the effective spin Hamiltonian reveal a two-stage phase ordering of the SkL phase after a thermal quench. The initial fast crystallization of skyrmions is followed by a slow relaxation dominated by the annihilation dynamics of dislocations, which are topological defects of the constituent magnetic stripe orders. The late-stage phase ordering also exhibits a dynamical scaling symmetry. We further show that the annihilation of dislocations follows a power-law time dependence with a logarithmic correction that depends on magnetic fields. Implications of our results for SkL phases in magnetic materials are also discussed. 
\end{abstract}


\maketitle

Complex magnetic textures such as vortices and skyrmions are not only of great fundamental interest in magnetism but also have important  implications in the emerging technology of spintronics~\cite{Nagaosa2013, Everschor2018, Back2020, Gobel2021}. 
Both vortices and skyrmions are nano-sized particle-like spin-textures characterized by nontrivial topological invariants. 
The presence of such complex patterns in metallic magnets could give rise to intriguing electronic and transport properties due to a nontrivial Berry phase acquired by electrons when traversing over closed loops of noncollinear or noncoplanar spins~\cite{Volovik1987,Berry1984,Nagaosa2012-1,Nagaosa2013}. The well-studied topological Hall effects~\cite{Loss1992,Ye1999,Bruno2004,Onoda2004,Binz2008,Nakazawa2019} and topological Nernst effects~\cite{Shiomi2013,Mizuta2016,Hirschberger2020TNE} are some of the representative examples. Also importantly, such topological electronic responses in metallic magnets can be controlled via the manipulation of magnetic textures. 

In magnetic materials, skyrmions are often stabilized  in the form of a skyrmion lattice (SkL), which is a periodic array of such particle-like topological spin-textures. Indeed, SkL as a spontaneous ground state was already predicted in the pioneering work of Bogdanov and Yablonskii that later triggered the enormous interest in magnetic skyrmions~\cite{Bogdanov1989,Roessler2006}. SkLs have since been reported in several chiral magnets such as MnSi~\cite{Muhlbauer2009} and other B20 compounds, as well as centrosymmetric materials~\cite{Tokura2020}. While the picture of SkL as an array of particle-like objects offers an intuitive framework to understand certain structural and dynamical aspects of skyrmion phases~\cite{Reichhardt2022}, recent studies have revealed the close connection between SkL and multiple-$Q$ magnetic ordering driven by partial nesting of the electron Fermi surface~\cite{Ozawa2016,Hayami2017,Hayami2021topological}. Indeed, this mechanism has been conjectured to be ubiquitous for itinerant spin systems with a wide range of filling fractions. 


Despite the huge interest and extensive research on magnetic SkL over the past decades, the phase-ordering dynamics of 
SkLs remains an open subject. Specifically, here one concerns the dynamical evolution and potential universal behaviors of skyrmion crystallization when a magnet is quenched into a skyrmion phase. It is known that the kinetics of phase ordering depends crucially on topological defects of the symmetry-breaking phase. Several super-universality classes of domain-growth laws have been established over the years~\cite{Bray1994,Puri2009}. The fact that both magnetic and translational symmetries are broken in a skyrmion crystal indicates rich relaxational dynamics of SkL phases which has yet to be systematically investigated. Understanding the phase ordering of SkL is also crucial to the engineering and control of skyrmion phases in real materials.

In this paper, we make an important step toward this goal by investigating the crystallization dynamics of a SkL in a realistic model of chiral metallic magnets. A minimum microscopic model for such itinerant spin systems is the s-d  Hamiltonian $\mathcal{H}_{\rm sd} = \sum_{\mathbf k, \sigma} \epsilon_{\mathbf k, \sigma} c^\dagger_{\mathbf k, \sigma} c^{\,}_{\mathbf k, \sigma} - J_{\rm sd} \sum_i \mathbf S_i \cdot c^\dagger_{i\sigma} \bm \sigma_{\sigma, \sigma'} c^{\,}_{i \sigma'}$, where the first term describes electron hopping on a lattice with $t_{ij}$ the transfer integrals and the second term represents a local coupling between the itinerant s-electron and the local moments $\mathbf S_i$ of d-electrons; $J_{\rm sd}$ represents the coupling strength. Spin-orbit coupling (SOC) within such single-band models can be described by either Rashba or Dresselhaus hopping terms. Dynamical simulations based on such electron models requires solving a disordered electron tight-binding Hamiltonian at every time-step, which could be prohibitively expensive for large systems. Instead, here we consider an effective spin Hamiltonian, similar in spirit to the Ruderman-Kittel-Kasuya-Yosida (RKKY) interaction~\cite{Ruderman1954,Kasuya1956,Yosida1957}, with SOC properly included~\cite{Hayami2018}.  At the leading second-order perturbation, the effective Hamiltonian has the  general form 
\begin{eqnarray}
	\mathcal{H} = -\sum_{ij} \mathbf S_i \cdot \mathsf{J}(\mathbf r_i - \mathbf r_j) \cdot \mathbf S_j - \sum_i \mathbf H \cdot \mathbf S_i.
\label{eq:Hamiltonian}
\end{eqnarray}
Here we have included a Zeeman coupling to an external field ${\bf H}=(0,0,H)$, and $\mathsf{J}(\mathbf r)$ represents an effective $3\times 3$ interaction matrix between two spins separated by $\mathbf r$. Its Fourier transform is given by 
\begin{eqnarray}
	\tilde{\mathsf{J}}(\mathbf q) =\frac{J_{\rm sd}^2}{N}  \sum_{\mathbf k} \sum_{\sigma,\sigma'} \chi^0_{\sigma\sigma'}(\mathbf k, \mathbf q)
	\left[ \mathsf{I} +  \mathsf{F}^{\sigma\sigma'}(\mathbf k, \mathbf q) \right],
\end{eqnarray} 
where $\chi^0_{\sigma\sigma'} = [f(\epsilon_{\mathbf k, \sigma}) - f(\epsilon_{\mathbf k + \mathbf q, \sigma'})]/(\epsilon_{\mathbf k + \mathbf q, \sigma'} - \epsilon_{\mathbf k, \sigma})$ is the spin-dependent susceptibility, $\epsilon_{\mathbf k, \sigma}$ is the electron band energy, $f(\epsilon)$ is the Fermi-Dirac function, $\mathsf{I}$ is the identity matrix, and the dimensionless matrix $\mathsf{F}^{\sigma\sigma'}$ accounts for the anisotropic spin interaction due to SOC~\cite{Hayami2018,1_supp_Heff}.

The interaction matrix $\tilde{\mathsf{J}}(\mathbf q)$ is often dominated by a few wave vectors $\mathbf Q_\eta$ when part of the electron Fermi surface is connected by them, i.e. $\epsilon_{\mathbf k + \mathbf Q_\eta, \sigma'} \approx \epsilon_{\mathbf k, \sigma}$. Such partial nesting of the Fermi surface has been shown to be a primary mechanism for the stabilization of skyrmion or vortex lattices in metallic magnets~\cite{Hayami2017}. An effective real-space Hamiltonian is then given by the inverse Fourier transformation of $\tilde{\mathsf{J}}(\mathbf q)$, which likely can only be done numerically for general dispersion relation $\epsilon_{\mathbf k, \sigma}$. 
As a first-order approximation, effective spin Hamiltonians can be obtained by keeping only contribution from the nesting wave vectors: $\tilde{\mathsf{J}}(\mathbf q) \approx \sum_{\eta} \tilde{\mathsf{J}}(\mathbf Q_\eta) \delta(\mathbf q - \mathbf Q_\eta)$. This approach has been employed to investigate complex spin textures in the ground state of itinerant magnets~\cite{Hayami2017,Hayami2021topological,Okumura2020,Yambe2021skyrmion,Shimizu2021anisotropy}. 
However, this approach gives rise to unrealistic infinite-range spin interactions. A more realistic analytical approach, while preserving the correct form of the anisotropic interaction, is to replace the $\delta$-function by a peak-shape function $h(\mathbf q)$ of a finite width, giving rise to spin-spin interactions which decay with distance in real space~\cite{Kato2021}. 

For concreteness, here we apply the procedure described above to a square-lattice s-d model with a Rashba SOC. A square array of skyrmions can be stabilized by partial nesting with two wave vectors $\mathbf Q_1 = (Q, 0)$ and $\mathbf Q_2 = (0, Q)$. The resultant real-space interaction matrix is given by
\begin{eqnarray}
	\label{eq:J_r}
	\mathsf{J}(\mathbf r) = \sum_{\eta = 1, 2} g(\mathbf r) \left[ {\rm Re}\mathsf{J}_\eta \cos(\mathbf Q_\eta\! \cdot\! \mathbf r) - {\rm Im}\mathsf{J}_\eta \sin(\mathbf Q_\eta \!\cdot\! \mathbf r) \right], \quad
\end{eqnarray}
where
\begin{eqnarray}
	\mathsf{J}_1 = \left(\begin{array}{ccc} J^{\perp} & 0 & 0 \\ 0 & J^{\perp} & -i D \\ 0 & i D & J^{zz} \end{array} \right)\!, \,\,
	\mathsf{J}_2 = \left(\begin{array}{ccc} J^{\perp} & 0 & -i D \\ 0 & J^{\perp} & 0 \\ i D & 0 & J^{zz} \end{array} \right)\!. \quad
\end{eqnarray}
and $J^\perp$, $J^{zz}$ describe electron-mediated exchange interactions, and $D$ represents effective long-ranged Dzyaloshinskii-Moriya interaction (DMI) induced by SOC; both $J$ and $D$ are of the order of $J_{\rm sd}^2/W$, where $W$ is the electron bandwidth. In the following, we set $2J^\perp + J^z = 1$ to serve as the unit for energy (and inverse time) and $D=0.3$. 
The function $g(\mathbf r)$ describes the decay of spin-spin interaction with distance. For simplicity, we assume a Lorentzian function for $h(\mathbf q)$, which leads to an exponential decaying $g(\mathbf r) = \mathcal{A}\, e^{-\gamma(|x| + |y|)}$, where $\mathcal{A}$ is a normalization constant~\cite{1_supp_Heff}. A hard cutoff $r_c$ such that $\mathsf{J}(\mathbf r) = 0$ for $|x|+|y| > r_c$ is further introduced for large-scale simulations of $N=1000^2$ spins. For results presented in the following, parameters  $\gamma=0.3$ and $r_{\rm c}=16$ are used.

\begin{figure}[tb]
\centering
\includegraphics[width=1.0\columnwidth]{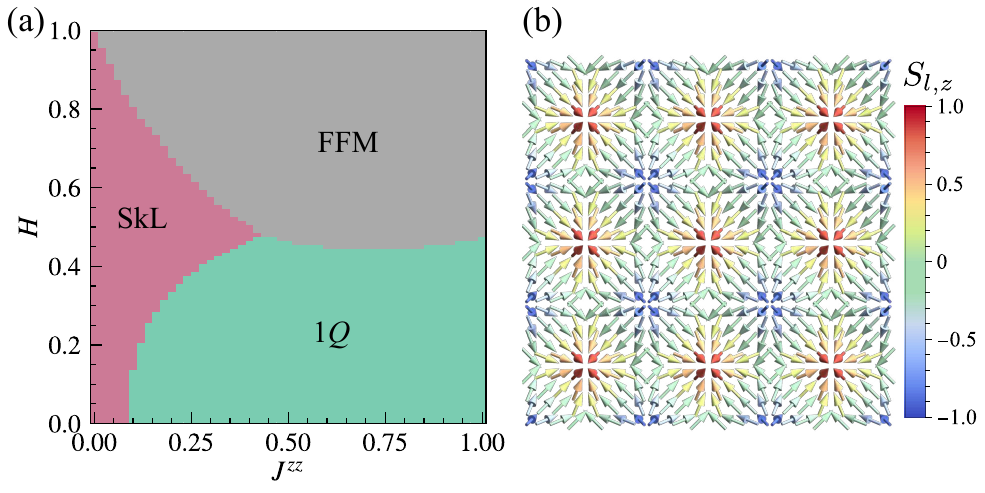}
\caption{
\label{fig:fig1}
(a) The ground state phase diagram includes a square Skyrmion lattice (SkL), an 1$Q$ state, and a forced ferromagnetic state (FFM).
(b) The real-space spin texture for $J^{zz}=0$ and $H=0$. The color of the arrows represents the out-of-plane component of spins.
}
\end{figure}

The dynamical evolution of the magnet is described by the Landau-Lifshitz-Gilbert (LLG) equation
\begin{eqnarray}
	\frac{d\mathbf S_i}{dt} = \frac{1}{1+\alpha^2}\left[ \frac{\partial \mathcal{H}}{\partial \mathbf S_i} \times \mathbf S_i
	+ \alpha \mathbf S_i \times \left(\mathbf S_i \times \frac{\partial \mathcal{H}}{\partial \mathbf S_i}\right) \right],
\end{eqnarray}
where $\alpha$ is the Gilbert damping coefficient, which is set to 0.1 in the following simulations. A fourth-order Runge-Kutta method is used to integrate the LLG equation with a time-step $\Delta t = 0.05$. 
The phase diagram, shown in Fig.~\ref{fig:fig1}(a) in the plane of exchange anisotropy $J^{zz}$ versus field $H$, includes a 1$Q$-cycloidal order, a double-$Q$ SkL, and a forced ferromagnetic state at high field. It is worth noting that our results are consistent with the phase diagram obtained from simulated annealing minimization of the infinite-range effective spin model~\cite{Hayami2018}. The double-$Q$ magnetic order at $H = 0$, shown in Fig.~\ref{fig:fig1}(b), exhibits a non-coplanar N\'eel-type vortex texture.


\begin{figure}[tb]
\centering
\includegraphics[width=1.0\columnwidth]{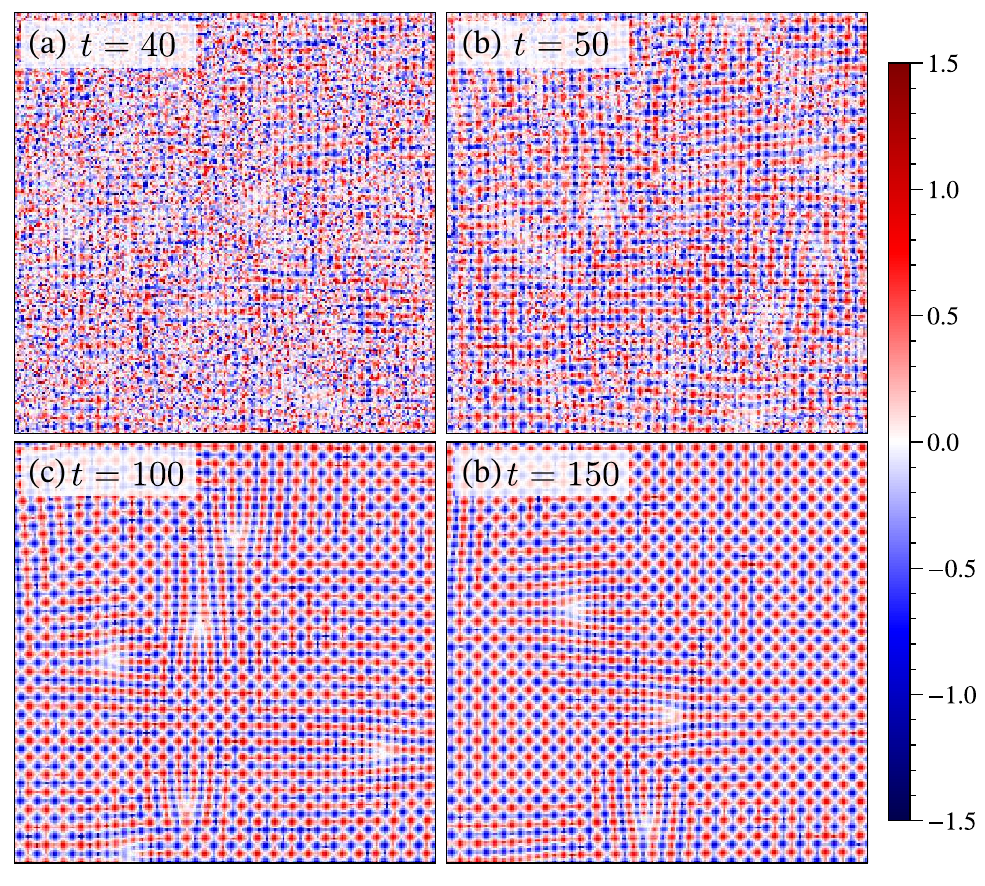}
\caption{
\label{fig:fig2}
Spatial distribution of the scalar spin chirality $\chi_i^{\rm sc}
$ obtained from spin configurations at different times after a thermal quench. Parameters $J^{ zz}=0$, and $H=0$ are used in the LLG simulation of a $1000^2$ system. Shown here are a selected $200\times 200$ region of the lattice.  
}
\end{figure}

Importantly, for small $J^{zz}$, the SkL is stable for a wide range of magnetic field, allowing us to study the crystallization dynamics of skyrmions and the field effects. To this end, the LLG dynamics was employed to simulate thermal quenches of a spin system with the effective RKKY interaction in Eq.~(\ref{eq:J_r}). An initial state of random spins, corresponding to equilibrium at high temperatures, is suddenly quenched to zero temperature at time $t = 0$. A typical example of the subsequent evolution of spins is shown in Fig.~\ref{fig:fig2} for the case of zero-field quench. The color shows the scalar chirality which is defined as 
$\chi_i^{\rm sc} = \sum_{\triangle_i} \mathbf S_1 \cdot \mathbf S_2 \times \mathbf S_3/2$, 
where the summation is taken over four triplets of nearest spins on a triangle. 
As skyrmions, including the N\'eel vortex texture at zero field, are characterized by non-coplanar spins that wrap around a sphere, the emergence of a SkL domain is indicated by the staggered checkerboard-like arrangement of positive and negative scalar chirality. As shown in Figs.~\ref{fig:fig2}(a) and (b), small patches of skyrmion arrays quickly emerge after the thermal quench. However, long-range coherence is yet to be established between different patches, and large areas of vanishing scalar chirality mark the boundaries between skyrmion crystallites.

As long-range crystallization order further develops, the incoherent regions quickly contract to particle-like objects of similar size as a skyrmion as shown in Figs.~\ref{fig:fig2}(c) and (d). These ``particles'', also characterized by a vanishing $\chi_i^{\rm sc}$, correspond to dislocation defects, which are topological defects associated with broken translational symmetries. The dislocations here correspond to a starting point of an extra row or column of skyrmion lines in the square lattice. These particle-like defects carry a topological charge corresponding to the so-called Burgers vector. Their topological nature also manifests itself in the fact that dislocations are created and annihilated in pairs. The phase-ordering of SkL is now dominated by the dynamics of dislocation defects.

\begin{figure}[tb]
\centering
\includegraphics[width=1.0\columnwidth]{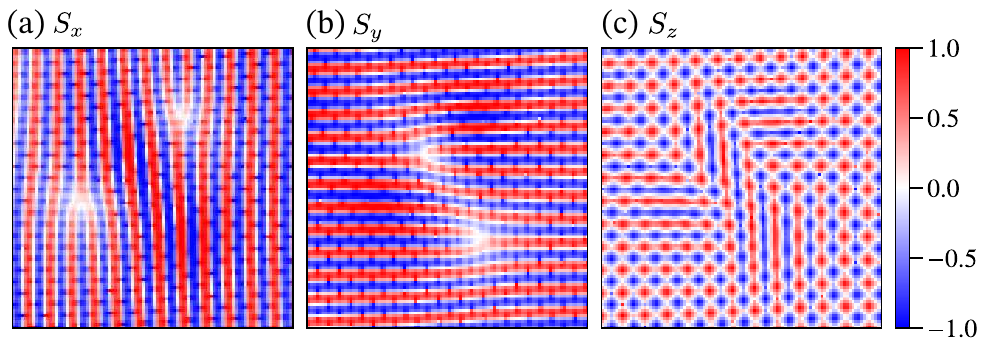}
\caption{
\label{fig:fig3}
Spatial profile of the three components of spin field $\mathbf S_i
$ at late stage of the phase ordering of SkL. 
Parameters are the same as those in Fig.~\ref{fig:fig2}.
}
\end{figure}

The magnetic structure of the double-$Q$ SkL can be approximated by an equal superposition of two spirals: 
\begin{eqnarray}
\mathbf S(\mathbf r_i) \sim 
\left( 
\cos\mathcal{Q}_1 , 
\cos\mathcal{Q}_2 , 
b (\sin\mathcal{Q}_1 + \sin\mathcal{Q}_2 )+ m_0 
\right), \quad
\label{eq:SkLansatz}
\end{eqnarray}
where $\mathcal{Q}_\eta = \mathbf Q_\eta \cdot \mathbf r_i +$ const, and the coefficients $b$ and $m_0$ depend on model parameters and magnetic field. Importantly, the $x$- and $y$-component of the spin field correspond to simple unidirectional stripes along the $x$ and $y$ directions, respectively. The fundamental defects of stripe order are also dislocations, as shown in Figs.~\ref{fig:fig3}(a) and (b). We can thus further classify the dislocation defects of SkL according to whether it is associated with the $S_x$ or $S_y$ component. Indeed, our simulations find that pair annihilations are possible only for dislocations of the same spin component. 

\begin{figure}[tb]
\centering
\includegraphics[width=1.0\columnwidth]{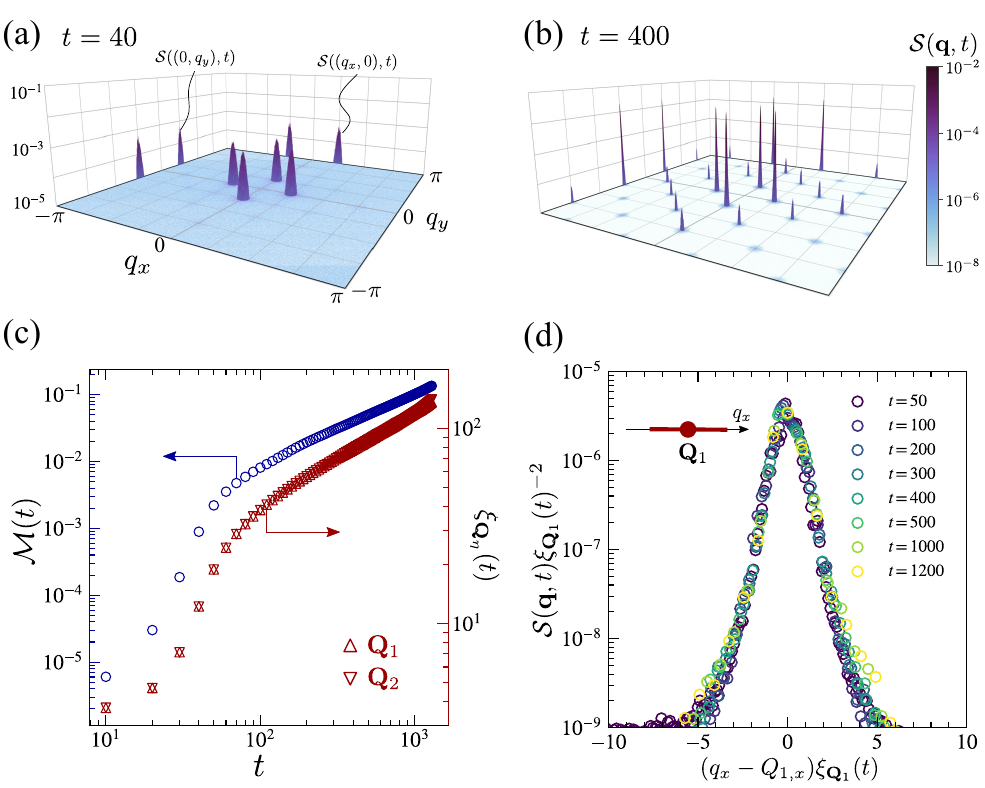}
\caption{
\label{fig:fig4}
Spin structure factor $\mathcal{S}(\mathbf q, t)$ with $J^{\rm zz}=0$ and $H=0$ at (a) $t=40$ and (b) $t=400$. 
(c) The SkL order parameter $\mathcal{M}(t) = \mathcal{S}({\bf Q}_1,t)+\mathcal{S}({\bf Q}_2,t)$ and the correlation length $\xi_{{\bf Q}_{\eta}}$ versus time. Both exhibit a power-law growth $t^{\alpha}$ with exponent $\alpha_{\mathcal{M}} \approx 1.046$ and $\alpha_{\xi} \approx 0.476$, respectively. 
(d)~Scaling plot of the structure factor around the peak at ${\bf Q}_1$ along the $q_x$ axis. 
}
\end{figure}

To further quantify the phase ordering of SkL, we examine the time-dependent spin structure factor, defined as $\mathcal{S}(\mathbf q, t) = \frac{1}{N^2}\braket{\left| \sum_i \mathbf S_i(t) \exp(i \mathbf q \cdot \mathbf r_i) \right|^2}$, where $\braket{\cdots}$ denotes the average of independent initial conditions. Examples of the structure factor at the early and late stages of phase ordering are shown in Figs.~\ref{fig:fig4}(a) and (b), respectively. 
The structure factor exhibits four broad peaks  at $\pm{\bf Q}_1$ and $\pm{\bf Q}_2$, where ${\bf Q}_1=(Q,0)$, ${\bf Q}_2=(0, Q)$, and $Q=0.785$, quickly after the quench; see e.g. Fig.~\ref{fig:fig4}(a) for  $t = 40$. 
These peaks at the nesting wave vectors become sharper as the system relaxes toward equilibrium, as shown in Fig.~\ref{fig:fig4}(b). Moreover, satellite peaks at $n_1{\bf Q}_1+n_2{\bf Q}_2$ with $n_1+n_2=2n+1$ ($n$ is an integer) start to emerge at late times, signaling the onset of higher harmonics of the constituent spiral orders.

An overall order parameter of the SkL phase can be defined as the sum of peak intensities $\mathcal{M}(t) = \mathcal{S}({\bf Q}_1,t)+\mathcal{S}({\bf Q}_2,t)$. As shown in Fig.~\ref{fig:fig4}(c), this SkL order parameter clearly exhibits a two-stage ordering discussed above: the fast development of quasi-long-ranged crystalline domains ($t \lesssim 80$), followed by a slow power-law growth dominated by the annihilation dynamics of dislocations ($t \gtrsim 80$). To characterize the late-stage phase ordering, Fig.~\ref{fig:fig4}(c) also shows the growth of the correlation length defined as the inverse widths of the Fourier peaks: $\xi_{{\bf Q}_{\eta}}=1/\Delta Q_{\eta}$, where $\Delta Q_{\eta}=\sum_{{\bf q}\sim{\bf Q}_{\eta}}|{\bf q}-{\bf Q}_{\eta}|\mathcal{S}(\mathbf q, t)/\sum_{{\bf q}\sim{\bf Q}_{\eta}}\mathcal{S}(\mathbf q, t)$. 
Furthermore, the power-law growth of the correlation lengths is intimately related to that of SkL order parameter $\mathcal{M}$. Indeed, we find that the late-stage ordering of SkL exhibits a dynamical symmetry. The structure factor at different times can be described by a universal function $\mathcal{F}$ with proper rescaling
\begin{eqnarray}
	\mathcal{S}(\mathbf q, t)= \xi_{\mathbf Q_\eta}^{2}\!(t) \, \mathcal{F}\!\left(|\mathbf q - \mathbf Q_\eta| \, \xi^{\,}_{\mathbf Q_\eta}(t) \right).
\end{eqnarray}
This is illustrated by the excellent data points collapsing shown in Fig.~\ref{fig:fig4}(d).


\begin{figure}[tb]
\centering
\includegraphics[width=0.9\columnwidth]{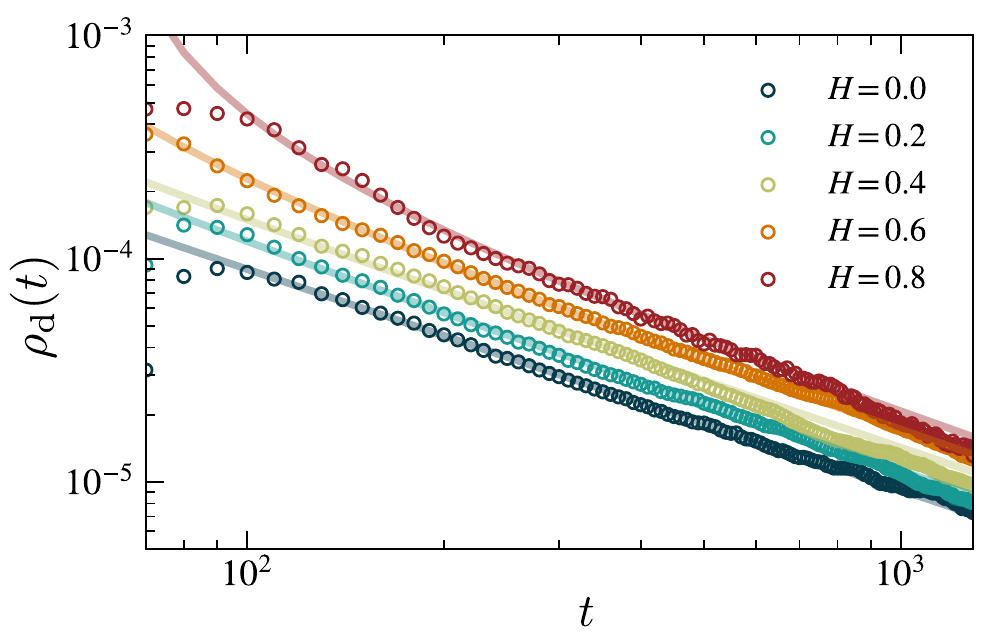}
\caption{
\label{fig:fig5}
Real-time dependence of the density of dislocations defects $\rho_{\rm d}(t)$ with $J^{zz}=0$ for varying magnetic field $H$, obtained by averaging over 32 independent runs.  The positions of defects are detected from the energy density~\cite{2_supp_dislocation}.
The pale lines represent the fitting by the formula $\rho_{\rm d}(t)/(\log(\rho_0/\rho_{\rm d}(t)) - 1) = a (t - t_0)^{-1}$. 
}
\end{figure}

The above late-time power-law behaviors are related to the dynamics of dislocations, which are topological defects of emergent square skyrmion arrays discussed above. As the long-range coherence of a crystalline order is disrupted by dislocations, the correlation length of SkL can be interpreted as the average distance $\ell$ between dislocations, which is related to their number density as $\ell \sim \rho_{\rm d}^{-1/2}$. The power-law growth of correlation length $\xi \sim t^{\alpha}$ thus implies a power-law decrease of dislocation density
$\rho_{\rm d} \sim t^{-\eta}$ 
with the exponent $\eta = 2 \alpha$. This is indeed confirmed in our large-scale LLG simulations summarized in Fig.~\ref{fig:fig5}, where a naive power-law fitting gives an exponent $\eta$ that depends weakly on magnetic field $H$. For example, the zero-field exponent $\eta \sim 1.03$ is consistent with the grow exponent of the correlation length shown in Fig.~\ref{fig:fig4}(c).

To understand the power-law annihilation of dislocations, we note that these topological defects can be mapped to vortices of effective 2D-XY models. As discussed in Eq.~(\ref{eq:SkLansatz}), the SkL can be viewed as comprised of two spiral spin orders. The $x$ and $y$ components of the coarse-grained spin field represent two orthogonal stripes
$S_x(\mathbf r) \sim \cos[\mathbf Q_1\! \cdot \! \mathbf r + \theta_1(\mathbf r)]$ and 
$S_y(\mathbf r) \sim  \cos[\mathbf Q_2\! \cdot \!\mathbf r + \theta_2(\mathbf r)]$. The dislocations associated with the two stripes, shown in Figs.~\ref{fig:fig3}(a) and (b), correspond to vortex singularities of the phase fields~$\theta_{1, 2}(\mathbf r)$. At the leading-order approximation, our system can thus be described by two coupled XY models: 
\begin{eqnarray}
	\mathcal{E} = A\int  \{ [(\nabla\theta_1)^2 + (\nabla\theta_2)^2] + 2\kappa (\nabla\theta_1 \cdot \nabla\theta_2) \} d^2\mathbf r,
\end{eqnarray}
where $A > 0$ represents the stiffness of the phase fields, and $\kappa$ denotes their coupling. A large $\kappa$ could induce a bound state of vortices from the two XY fields. Nonetheless, it is straightforward to show that the above action is equivalent to two independent XY fields $\theta_{\pm} = \theta_1 \pm \theta_2$ with different stiffness: $A_{\pm} = A(1\pm \kappa)/2$.
Importantly, it has been shown that the phase ordering of the 2D XY model is governed by the annihilation of vortices, whose number density follows a power-law $\rho_{\rm v} \sim t^{-1}$ time dependence, up to a logarithmic correction~\cite{Yurke1993,Bray2000}. Our results thus strongly indicate that the ordering kinetics of the SkL phase belongs to the same dynamical universality class of 2D XY model. In fact, while the extracted exponent is weakly dependent on $H$, the fact that all $\eta$ values lie in the vicinity of unity suggests a universal $\rho_{\rm d} \sim t^{-1}$ behavior, yet with a field-dependent logarithmic correction. We have checked that the various curves in Fig.~\ref{fig:fig5} indeed can be well described by the formula $\rho_{\rm d}/(\log(\rho_0/\rho_{\rm d}) - 1) = a (t - t_0)^{-1}$, where $\rho_0$, $t_0$, and $a$ are fitting parameters~\cite{Yurke1993}.

To conclude, we have presented a comprehensive study on the ordering dynamics of SkL phases in chiral metallic magnets, which is important to the design and engineering of skyrmion-based spintronics devices. Fundamentally, magnetic skyrmion lattices also provide another platform to study crystallization phenomena in two dimensions, a field which has yet to be systematically investigated. An intriguing fast crystallization process has recently been observed in the ``square ice'' formed by water molecules locked between two graphene sheets~\cite{Siller2015}. While our work here sheds new light on this fundamental subject of 2D crystallization dynamics, several important issues remain to be addressed. For example, although the long-range nature of electron-mediated interactions might be crucial to the initial fast crystallization, a detailed study on the effect of interaction range and its interplay with other factors is desired. The built-in chirality due to SOC is another crucial component for the fast establishment of coherent lattices. The recent experimental observation of square SkL in centrosymmetric magnets such as GdRu$_2$Si$_2$~\cite{Khanh2020} and EuAl$_4$~\cite{Takagi2022} thus calls for theoretical studies of skyrmion crystallization dynamics in non-chiral itinerant magnets~\cite{Hayami2017, Hayami2021square}. Yet another interesting question is the effects of crystal geometry. More complicated annihilation dynamics of dislocations might arise in the triangular SkL observed in many skyrmion materials.

\begin{acknowledgments}
The authors thank Y. Kato and Y. Motome for fruitful discussions. 
This work was supported by JSPS KAKENHI Grant Number No. JP21J20812. 
K.S. was supported by the Program for Leading Graduate Schools (MERIT-WINGS). Parts of the numerical calculations were performed in the supercomputing systems in ISSP, the University of Tokyo. G.W.C was partially supported by the US Department of Energy Basic Energy Sciences under Contract No. DE-SC0020330.
\end{acknowledgments}

\bibliography{ref}

\begin{thebibliography}{42}%
\makeatletter
\providecommand \@ifxundefined [1]{%
 \@ifx{#1\undefined}
}%
\providecommand \@ifnum [1]{%
 \ifnum #1\expandafter \@firstoftwo
 \else \expandafter \@secondoftwo
 \fi
}%
\providecommand \@ifx [1]{%
 \ifx #1\expandafter \@firstoftwo
 \else \expandafter \@secondoftwo
 \fi
}%
\providecommand \natexlab [1]{#1}%
\providecommand \enquote  [1]{``#1''}%
\providecommand \bibnamefont  [1]{#1}%
\providecommand \bibfnamefont [1]{#1}%
\providecommand \citenamefont [1]{#1}%
\providecommand \href@noop [0]{\@secondoftwo}%
\providecommand \href [0]{\begingroup \@sanitize@url \@href}%
\providecommand \@href[1]{\@@startlink{#1}\@@href}%
\providecommand \@@href[1]{\endgroup#1\@@endlink}%
\providecommand \@sanitize@url [0]{\catcode `\\12\catcode `\$12\catcode
  `\&12\catcode `\#12\catcode `\^12\catcode `\_12\catcode `\%12\relax}%
\providecommand \@@startlink[1]{}%
\providecommand \@@endlink[0]{}%
\providecommand \url  [0]{\begingroup\@sanitize@url \@url }%
\providecommand \@url [1]{\endgroup\@href {#1}{\urlprefix }}%
\providecommand \urlprefix  [0]{URL }%
\providecommand \Eprint [0]{\href }%
\providecommand \doibase [0]{https://doi.org/}%
\providecommand \selectlanguage [0]{\@gobble}%
\providecommand \bibinfo  [0]{\@secondoftwo}%
\providecommand \bibfield  [0]{\@secondoftwo}%
\providecommand \translation [1]{[#1]}%
\providecommand \BibitemOpen [0]{}%
\providecommand \bibitemStop [0]{}%
\providecommand \bibitemNoStop [0]{.\EOS\space}%
\providecommand \EOS [0]{\spacefactor3000\relax}%
\providecommand \BibitemShut  [1]{\csname bibitem#1\endcsname}%
\let\auto@bib@innerbib\@empty
\bibitem [{\citenamefont {Nagaosa}\ and\ \citenamefont
  {Tokura}(2013)}]{Nagaosa2013}%
  \BibitemOpen
  \bibfield  {author} {\bibinfo {author} {\bibfnamefont {N.}~\bibnamefont
  {Nagaosa}}\ and\ \bibinfo {author} {\bibfnamefont {Y.}~\bibnamefont
  {Tokura}},\ }\bibfield  {title} {\bibinfo {title} {{Topological properties
  and dynamics of magnetic skyrmions}},\ }\href
  {https://doi.org/10.1038/nnano.2013.243} {\bibfield  {journal} {\bibinfo
  {journal} {Nat. Nanotechnol.}\ }\textbf {\bibinfo {volume} {8}},\ \bibinfo
  {pages} {899} (\bibinfo {year} {2013})}\BibitemShut {NoStop}%
\bibitem [{\citenamefont {Everschor-Sitte}\ \emph {et~al.}(2018)\citenamefont
  {Everschor-Sitte}, \citenamefont {Masell}, \citenamefont {Reeve},\ and\
  \citenamefont {Kläui}}]{Everschor2018}%
  \BibitemOpen
  \bibfield  {author} {\bibinfo {author} {\bibfnamefont {K.}~\bibnamefont
  {Everschor-Sitte}}, \bibinfo {author} {\bibfnamefont {J.}~\bibnamefont
  {Masell}}, \bibinfo {author} {\bibfnamefont {R.~M.}\ \bibnamefont {Reeve}},\
  and\ \bibinfo {author} {\bibfnamefont {M.}~\bibnamefont {Kläui}},\
  }\bibfield  {title} {\bibinfo {title} {{Perspective: Magnetic
  skyrmions—Overview of recent progress in an active research field}},\
  }\bibfield  {journal} {\bibinfo  {journal} {Journal of Applied Physics}\
  }\textbf {\bibinfo {volume} {124}},\ \href
  {https://doi.org/10.1063/1.5048972} {10.1063/1.5048972} (\bibinfo {year}
  {2018}),\ \bibinfo {note} {240901}\BibitemShut {NoStop}%
\bibitem [{\citenamefont {Back}\ \emph {et~al.}(2020)\citenamefont {Back},
  \citenamefont {Cros}, \citenamefont {Ebert}, \citenamefont {Everschor-Sitte},
  \citenamefont {Fert}, \citenamefont {Garst}, \citenamefont {Ma},
  \citenamefont {Mankovsky}, \citenamefont {Monchesky}, \citenamefont
  {Mostovoy}, \citenamefont {Nagaosa}, \citenamefont {Parkin}, \citenamefont
  {Pfleiderer}, \citenamefont {Reyren}, \citenamefont {Rosch}, \citenamefont
  {Taguchi}, \citenamefont {Tokura}, \citenamefont {von Bergmann},\ and\
  \citenamefont {Zang}}]{Back2020}%
  \BibitemOpen
  \bibfield  {author} {\bibinfo {author} {\bibfnamefont {C.}~\bibnamefont
  {Back}}, \bibinfo {author} {\bibfnamefont {V.}~\bibnamefont {Cros}}, \bibinfo
  {author} {\bibfnamefont {H.}~\bibnamefont {Ebert}}, \bibinfo {author}
  {\bibfnamefont {K.}~\bibnamefont {Everschor-Sitte}}, \bibinfo {author}
  {\bibfnamefont {A.}~\bibnamefont {Fert}}, \bibinfo {author} {\bibfnamefont
  {M.}~\bibnamefont {Garst}}, \bibinfo {author} {\bibfnamefont
  {T.}~\bibnamefont {Ma}}, \bibinfo {author} {\bibfnamefont {S.}~\bibnamefont
  {Mankovsky}}, \bibinfo {author} {\bibfnamefont {T.~L.}\ \bibnamefont
  {Monchesky}}, \bibinfo {author} {\bibfnamefont {M.}~\bibnamefont {Mostovoy}},
  \bibinfo {author} {\bibfnamefont {N.}~\bibnamefont {Nagaosa}}, \bibinfo
  {author} {\bibfnamefont {S.~S.~P.}\ \bibnamefont {Parkin}}, \bibinfo {author}
  {\bibfnamefont {C.}~\bibnamefont {Pfleiderer}}, \bibinfo {author}
  {\bibfnamefont {N.}~\bibnamefont {Reyren}}, \bibinfo {author} {\bibfnamefont
  {A.}~\bibnamefont {Rosch}}, \bibinfo {author} {\bibfnamefont
  {Y.}~\bibnamefont {Taguchi}}, \bibinfo {author} {\bibfnamefont
  {Y.}~\bibnamefont {Tokura}}, \bibinfo {author} {\bibfnamefont
  {K.}~\bibnamefont {von Bergmann}},\ and\ \bibinfo {author} {\bibfnamefont
  {J.}~\bibnamefont {Zang}},\ }\bibfield  {title} {\bibinfo {title} {The 2020
  skyrmionics roadmap},\ }\href {https://doi.org/10.1088/1361-6463/ab8418}
  {\bibfield  {journal} {\bibinfo  {journal} {J. Phys. D: Appl. Phys.}\
  }\textbf {\bibinfo {volume} {53}},\ \bibinfo {pages} {363001} (\bibinfo
  {year} {2020})}\BibitemShut {NoStop}%
\bibitem [{\citenamefont {Göbel}\ \emph {et~al.}(2021)\citenamefont {Göbel},
  \citenamefont {Mertig},\ and\ \citenamefont {Tretiakov}}]{Gobel2021}%
  \BibitemOpen
  \bibfield  {author} {\bibinfo {author} {\bibfnamefont {B.}~\bibnamefont
  {Göbel}}, \bibinfo {author} {\bibfnamefont {I.}~\bibnamefont {Mertig}},\
  and\ \bibinfo {author} {\bibfnamefont {O.~A.}\ \bibnamefont {Tretiakov}},\
  }\bibfield  {title} {\bibinfo {title} {Beyond skyrmions: Review and
  perspectives of alternative magnetic quasiparticles},\ }\href
  {https://doi.org/https://doi.org/10.1016/j.physrep.2020.10.001} {\bibfield
  {journal} {\bibinfo  {journal} {Physics Reports}\ }\textbf {\bibinfo {volume}
  {895}},\ \bibinfo {pages} {1} (\bibinfo {year} {2021})},\ \bibinfo {note}
  {beyond skyrmions: Review and perspectives of alternative magnetic
  quasiparticles}\BibitemShut {NoStop}%
\bibitem [{\citenamefont {Volovik}(1987)}]{Volovik1987}%
  \BibitemOpen
  \bibfield  {author} {\bibinfo {author} {\bibfnamefont {G.~E.}\ \bibnamefont
  {Volovik}},\ }\bibfield  {title} {\bibinfo {title} {{Linear momentum in
  ferromagnets}},\ }\href {https://doi.org/10.1088/0022-3719/20/7/003}
  {\bibfield  {journal} {\bibinfo  {journal} {J. Phys. C Solid State Phys.}\
  }\textbf {\bibinfo {volume} {20}},\ \bibinfo {pages} {L83} (\bibinfo {year}
  {1987})}\BibitemShut {NoStop}%
\bibitem [{\citenamefont {Berry}(1984)}]{Berry1984}%
  \BibitemOpen
  \bibfield  {author} {\bibinfo {author} {\bibfnamefont {M.~V.}\ \bibnamefont
  {Berry}},\ }\bibfield  {title} {\bibinfo {title} {{Quantal phase factors
  accompanying adiabatic changes}},\ }\href
  {https://doi.org/https://doi.org/10.1098/rspa.1984.0023} {\bibfield
  {journal} {\bibinfo  {journal} {Proc. R. Soc. Lond. A}\ }\textbf {\bibinfo
  {volume} {392}},\ \bibinfo {pages} {45} (\bibinfo {year} {1984})}\BibitemShut
  {NoStop}%
\bibitem [{\citenamefont {Nagaosa}\ and\ \citenamefont
  {Tokura}(2012)}]{Nagaosa2012-1}%
  \BibitemOpen
  \bibfield  {author} {\bibinfo {author} {\bibfnamefont {N.}~\bibnamefont
  {Nagaosa}}\ and\ \bibinfo {author} {\bibfnamefont {Y.}~\bibnamefont
  {Tokura}},\ }\bibfield  {title} {\bibinfo {title} {{Emergent electromagnetism
  in solids}},\ }\href {https://doi.org/10.1088/0031-8949/2012/t146/014020}
  {\bibfield  {journal} {\bibinfo  {journal} {Phys. Scr.}\ }\textbf {\bibinfo
  {volume} {T146}},\ \bibinfo {pages} {14020} (\bibinfo {year}
  {2012})}\BibitemShut {NoStop}%
\bibitem [{\citenamefont {Loss}\ and\ \citenamefont
  {Goldbart}(1992)}]{Loss1992}%
  \BibitemOpen
  \bibfield  {author} {\bibinfo {author} {\bibfnamefont {D.}~\bibnamefont
  {Loss}}\ and\ \bibinfo {author} {\bibfnamefont {P.~M.}\ \bibnamefont
  {Goldbart}},\ }\bibfield  {title} {\bibinfo {title} {{Persistent currents
  from {B}erry's phase in mesoscopic systems}},\ }\href
  {https://doi.org/10.1103/PhysRevB.45.13544} {\bibfield  {journal} {\bibinfo
  {journal} {Phys. Rev. B}\ }\textbf {\bibinfo {volume} {45}},\ \bibinfo
  {pages} {13544} (\bibinfo {year} {1992})}\BibitemShut {NoStop}%
\bibitem [{\citenamefont {Ye}\ \emph {et~al.}(1999)\citenamefont {Ye},
  \citenamefont {Kim}, \citenamefont {Millis}, \citenamefont {Shraiman},
  \citenamefont {Majumdar},\ and\ \citenamefont {Te\ifmmode \check{s}\else
  \v{s}\fi{}anovi\ifmmode~\acute{c}\else \'{c}\fi{}}}]{Ye1999}%
  \BibitemOpen
  \bibfield  {author} {\bibinfo {author} {\bibfnamefont {J.}~\bibnamefont
  {Ye}}, \bibinfo {author} {\bibfnamefont {Y.~B.}\ \bibnamefont {Kim}},
  \bibinfo {author} {\bibfnamefont {A.~J.}\ \bibnamefont {Millis}}, \bibinfo
  {author} {\bibfnamefont {B.~I.}\ \bibnamefont {Shraiman}}, \bibinfo {author}
  {\bibfnamefont {P.}~\bibnamefont {Majumdar}},\ and\ \bibinfo {author}
  {\bibfnamefont {Z.}~\bibnamefont {Te\ifmmode \check{s}\else
  \v{s}\fi{}anovi\ifmmode~\acute{c}\else \'{c}\fi{}}},\ }\bibfield  {title}
  {\bibinfo {title} {{Berry Phase Theory of the Anomalous Hall Effect:
  Application to Colossal Magnetoresistance Manganites}},\ }\href
  {https://doi.org/10.1103/PhysRevLett.83.3737} {\bibfield  {journal} {\bibinfo
   {journal} {Phys. Rev. Lett.}\ }\textbf {\bibinfo {volume} {83}},\ \bibinfo
  {pages} {3737} (\bibinfo {year} {1999})}\BibitemShut {NoStop}%
\bibitem [{\citenamefont {Bruno}\ \emph {et~al.}(2004)\citenamefont {Bruno},
  \citenamefont {Dugaev},\ and\ \citenamefont {Taillefumier}}]{Bruno2004}%
  \BibitemOpen
  \bibfield  {author} {\bibinfo {author} {\bibfnamefont {P.}~\bibnamefont
  {Bruno}}, \bibinfo {author} {\bibfnamefont {V.~K.}\ \bibnamefont {Dugaev}},\
  and\ \bibinfo {author} {\bibfnamefont {M.}~\bibnamefont {Taillefumier}},\
  }\bibfield  {title} {\bibinfo {title} {{Topological Hall Effect and Berry
  Phase in Magnetic Nanostructures}},\ }\href
  {https://doi.org/10.1103/PhysRevLett.93.096806} {\bibfield  {journal}
  {\bibinfo  {journal} {Phys. Rev. Lett.}\ }\textbf {\bibinfo {volume} {93}},\
  \bibinfo {pages} {096806} (\bibinfo {year} {2004})}\BibitemShut {NoStop}%
\bibitem [{\citenamefont {Onoda}\ \emph {et~al.}(2004)\citenamefont {Onoda},
  \citenamefont {Tatara},\ and\ \citenamefont {Nagaosa}}]{Onoda2004}%
  \BibitemOpen
  \bibfield  {author} {\bibinfo {author} {\bibfnamefont {M.}~\bibnamefont
  {Onoda}}, \bibinfo {author} {\bibfnamefont {G.}~\bibnamefont {Tatara}},\ and\
  \bibinfo {author} {\bibfnamefont {N.}~\bibnamefont {Nagaosa}},\ }\bibfield
  {title} {\bibinfo {title} {{Anomalous Hall Effect and Skyrmion Number in Real
  and Momentum Spaces}},\ }\href {https://doi.org/10.1143/JPSJ.73.2624}
  {\bibfield  {journal} {\bibinfo  {journal} {J. Phys. Soc. Jpn.}\ }\textbf
  {\bibinfo {volume} {73}},\ \bibinfo {pages} {2624} (\bibinfo {year}
  {2004})}\BibitemShut {NoStop}%
\bibitem [{\citenamefont {Binz}\ and\ \citenamefont
  {Vishwanath}(2008)}]{Binz2008}%
  \BibitemOpen
  \bibfield  {author} {\bibinfo {author} {\bibfnamefont {B.}~\bibnamefont
  {Binz}}\ and\ \bibinfo {author} {\bibfnamefont {A.}~\bibnamefont
  {Vishwanath}},\ }\bibfield  {title} {\bibinfo {title} {{Chirality induced
  anomalous-Hall effect in helical spin crystals}},\ }\href
  {https://doi.org/https://doi.org/10.1016/j.physb.2007.10.136} {\bibfield
  {journal} {\bibinfo  {journal} {Physica B: Condensed Matter}\ }\textbf
  {\bibinfo {volume} {403}},\ \bibinfo {pages} {1336 } (\bibinfo {year}
  {2008})}\BibitemShut {NoStop}%
\bibitem [{\citenamefont {Nakazawa}\ and\ \citenamefont
  {Kohno}(2019)}]{Nakazawa2019}%
  \BibitemOpen
  \bibfield  {author} {\bibinfo {author} {\bibfnamefont {K.}~\bibnamefont
  {Nakazawa}}\ and\ \bibinfo {author} {\bibfnamefont {H.}~\bibnamefont
  {Kohno}},\ }\bibfield  {title} {\bibinfo {title} {Weak coupling theory of
  topological hall effect},\ }\href
  {https://doi.org/10.1103/PhysRevB.99.174425} {\bibfield  {journal} {\bibinfo
  {journal} {Phys. Rev. B}\ }\textbf {\bibinfo {volume} {99}},\ \bibinfo
  {pages} {174425} (\bibinfo {year} {2019})}\BibitemShut {NoStop}%
\bibitem [{\citenamefont {Shiomi}\ \emph {et~al.}(2013)\citenamefont {Shiomi},
  \citenamefont {Kanazawa}, \citenamefont {Shibata}, \citenamefont {Onose},\
  and\ \citenamefont {Tokura}}]{Shiomi2013}%
  \BibitemOpen
  \bibfield  {author} {\bibinfo {author} {\bibfnamefont {Y.}~\bibnamefont
  {Shiomi}}, \bibinfo {author} {\bibfnamefont {N.}~\bibnamefont {Kanazawa}},
  \bibinfo {author} {\bibfnamefont {K.}~\bibnamefont {Shibata}}, \bibinfo
  {author} {\bibfnamefont {Y.}~\bibnamefont {Onose}},\ and\ \bibinfo {author}
  {\bibfnamefont {Y.}~\bibnamefont {Tokura}},\ }\bibfield  {title} {\bibinfo
  {title} {{Topological Nernst effect in a three-dimensional skyrmion-lattice
  phase}},\ }\href {https://doi.org/10.1103/PhysRevB.88.064409} {\bibfield
  {journal} {\bibinfo  {journal} {Phys. Rev. B}\ }\textbf {\bibinfo {volume}
  {88}},\ \bibinfo {pages} {064409} (\bibinfo {year} {2013})}\BibitemShut
  {NoStop}%
\bibitem [{\citenamefont {Mizuta}\ and\ \citenamefont
  {Ishii}(2016)}]{Mizuta2016}%
  \BibitemOpen
  \bibfield  {author} {\bibinfo {author} {\bibfnamefont {Y.~P.}\ \bibnamefont
  {Mizuta}}\ and\ \bibinfo {author} {\bibfnamefont {F.}~\bibnamefont {Ishii}},\
  }\bibfield  {title} {\bibinfo {title} {{Large anomalous Nernst effect in a
  skyrmion crystal}},\ }\href
  {https://doi.org/https://doi.org/10.1038/srep28076} {\bibfield  {journal}
  {\bibinfo  {journal} {Sci. Rep.}\ }\textbf {\bibinfo {volume} {6}},\ \bibinfo
  {pages} {28076} (\bibinfo {year} {2016})}\BibitemShut {NoStop}%
\bibitem [{\citenamefont {Hirschberger}\ \emph {et~al.}(2020)\citenamefont
  {Hirschberger}, \citenamefont {Spitz}, \citenamefont {Nomoto}, \citenamefont
  {Kurumaji}, \citenamefont {Gao}, \citenamefont {Masell}, \citenamefont
  {Nakajima}, \citenamefont {Kikkawa}, \citenamefont {Yamasaki}, \citenamefont
  {Sagayama}, \citenamefont {Nakao}, \citenamefont {Taguchi}, \citenamefont
  {Arita}, \citenamefont {Arima},\ and\ \citenamefont
  {Tokura}}]{Hirschberger2020TNE}%
  \BibitemOpen
  \bibfield  {author} {\bibinfo {author} {\bibfnamefont {M.}~\bibnamefont
  {Hirschberger}}, \bibinfo {author} {\bibfnamefont {L.}~\bibnamefont {Spitz}},
  \bibinfo {author} {\bibfnamefont {T.}~\bibnamefont {Nomoto}}, \bibinfo
  {author} {\bibfnamefont {T.}~\bibnamefont {Kurumaji}}, \bibinfo {author}
  {\bibfnamefont {S.}~\bibnamefont {Gao}}, \bibinfo {author} {\bibfnamefont
  {J.}~\bibnamefont {Masell}}, \bibinfo {author} {\bibfnamefont
  {T.}~\bibnamefont {Nakajima}}, \bibinfo {author} {\bibfnamefont
  {A.}~\bibnamefont {Kikkawa}}, \bibinfo {author} {\bibfnamefont
  {Y.}~\bibnamefont {Yamasaki}}, \bibinfo {author} {\bibfnamefont
  {H.}~\bibnamefont {Sagayama}}, \bibinfo {author} {\bibfnamefont
  {H.}~\bibnamefont {Nakao}}, \bibinfo {author} {\bibfnamefont
  {Y.}~\bibnamefont {Taguchi}}, \bibinfo {author} {\bibfnamefont
  {R.}~\bibnamefont {Arita}}, \bibinfo {author} {\bibfnamefont {T.-h.}\
  \bibnamefont {Arima}},\ and\ \bibinfo {author} {\bibfnamefont
  {Y.}~\bibnamefont {Tokura}},\ }\bibfield  {title} {\bibinfo {title}
  {{Topological Nernst Effect of the Two-Dimensional Skyrmion Lattice}},\
  }\href {https://doi.org/10.1103/PhysRevLett.125.076602} {\bibfield  {journal}
  {\bibinfo  {journal} {Phys. Rev. Lett.}\ }\textbf {\bibinfo {volume} {125}},\
  \bibinfo {pages} {076602} (\bibinfo {year} {2020})}\BibitemShut {NoStop}%
\bibitem [{\citenamefont {Bogdanov}\ and\ \citenamefont
  {Yablonskii}(1989)}]{Bogdanov1989}%
  \BibitemOpen
  \bibfield  {author} {\bibinfo {author} {\bibfnamefont {A.~N.}\ \bibnamefont
  {Bogdanov}}\ and\ \bibinfo {author} {\bibfnamefont {D.}~\bibnamefont
  {Yablonskii}},\ }\bibfield  {title} {\bibinfo {title} {{Thermodynamically
  stable ``vortices'' in magnetically ordered crystals. The mixed state of
  magnets}},\ }\href@noop {} {\bibfield  {journal} {\bibinfo  {journal} {Zh.
  Eksp. Teor. Fiz}\ }\textbf {\bibinfo {volume} {95}},\ \bibinfo {pages} {178}
  (\bibinfo {year} {1989})}\BibitemShut {NoStop}%
\bibitem [{\citenamefont {Roessler}\ \emph {et~al.}(2006)\citenamefont
  {Roessler}, \citenamefont {Bogdanov},\ and\ \citenamefont
  {Pfleiderer}}]{Roessler2006}%
  \BibitemOpen
  \bibfield  {author} {\bibinfo {author} {\bibfnamefont {U.~K.}\ \bibnamefont
  {Roessler}}, \bibinfo {author} {\bibfnamefont {A.}~\bibnamefont {Bogdanov}},\
  and\ \bibinfo {author} {\bibfnamefont {C.}~\bibnamefont {Pfleiderer}},\
  }\bibfield  {title} {\bibinfo {title} {{Spontaneous skyrmion ground states in
  magnetic metals}},\ }\href
  {https://doi.org/https://doi.org/10.1038/nature05056} {\bibfield  {journal}
  {\bibinfo  {journal} {Nature}\ }\textbf {\bibinfo {volume} {442}},\ \bibinfo
  {pages} {797} (\bibinfo {year} {2006})}\BibitemShut {NoStop}%
\bibitem [{\citenamefont {M{\"{u}}hlbauer}\ \emph {et~al.}(2009)\citenamefont
  {M{\"{u}}hlbauer}, \citenamefont {Binz}, \citenamefont {Jonietz},
  \citenamefont {Pfleiderer}, \citenamefont {Rosch}, \citenamefont {Neubauer},
  \citenamefont {Georgii},\ and\ \citenamefont {B{\"{o}}ni}}]{Muhlbauer2009}%
  \BibitemOpen
  \bibfield  {author} {\bibinfo {author} {\bibfnamefont {S.}~\bibnamefont
  {M{\"{u}}hlbauer}}, \bibinfo {author} {\bibfnamefont {B.}~\bibnamefont
  {Binz}}, \bibinfo {author} {\bibfnamefont {F.}~\bibnamefont {Jonietz}},
  \bibinfo {author} {\bibfnamefont {C.}~\bibnamefont {Pfleiderer}}, \bibinfo
  {author} {\bibfnamefont {A.}~\bibnamefont {Rosch}}, \bibinfo {author}
  {\bibfnamefont {A.}~\bibnamefont {Neubauer}}, \bibinfo {author}
  {\bibfnamefont {R.}~\bibnamefont {Georgii}},\ and\ \bibinfo {author}
  {\bibfnamefont {P.}~\bibnamefont {B{\"{o}}ni}},\ }\bibfield  {title}
  {\bibinfo {title} {{Skyrmion Lattice in a Chiral Magnet}},\ }\href
  {https://doi.org/10.1126/science.1166767} {\bibfield  {journal} {\bibinfo
  {journal} {Science}\ }\textbf {\bibinfo {volume} {323}},\ \bibinfo {pages}
  {915} (\bibinfo {year} {2009})}\BibitemShut {NoStop}%
\bibitem [{\citenamefont {Tokura}\ and\ \citenamefont
  {Kanazawa}(2020)}]{Tokura2020}%
  \BibitemOpen
  \bibfield  {author} {\bibinfo {author} {\bibfnamefont {Y.}~\bibnamefont
  {Tokura}}\ and\ \bibinfo {author} {\bibfnamefont {N.}~\bibnamefont
  {Kanazawa}},\ }\bibfield  {title} {\bibinfo {title} {{Magnetic Skyrmion
  Materials}},\ }\href@noop {} {\bibfield  {journal} {\bibinfo  {journal}
  {Chem. Rev.}\ } (\bibinfo {year} {2020})}\BibitemShut {NoStop}%
\bibitem [{\citenamefont {Reichhardt}\ \emph {et~al.}(2022)\citenamefont
  {Reichhardt}, \citenamefont {Reichhardt},\ and\ \citenamefont {Milo\ifmmode
  \check{s}\else \v{s}\fi{}evi\ifmmode~\acute{c}\else
  \'{c}\fi{}}}]{Reichhardt2022}%
  \BibitemOpen
  \bibfield  {author} {\bibinfo {author} {\bibfnamefont {C.}~\bibnamefont
  {Reichhardt}}, \bibinfo {author} {\bibfnamefont {C.~J.~O.}\ \bibnamefont
  {Reichhardt}},\ and\ \bibinfo {author} {\bibfnamefont {M.~V.}\ \bibnamefont
  {Milo\ifmmode \check{s}\else \v{s}\fi{}evi\ifmmode~\acute{c}\else
  \'{c}\fi{}}},\ }\bibfield  {title} {\bibinfo {title} {Statics and dynamics of
  skyrmions interacting with disorder and nanostructures},\ }\href
  {https://doi.org/10.1103/RevModPhys.94.035005} {\bibfield  {journal}
  {\bibinfo  {journal} {Rev. Mod. Phys.}\ }\textbf {\bibinfo {volume} {94}},\
  \bibinfo {pages} {035005} (\bibinfo {year} {2022})}\BibitemShut {NoStop}%
\bibitem [{\citenamefont {Ozawa}\ \emph {et~al.}(2016)\citenamefont {Ozawa},
  \citenamefont {Hayami}, \citenamefont {Barros}, \citenamefont {Chern},
  \citenamefont {Motome},\ and\ \citenamefont {Batista}}]{Ozawa2016}%
  \BibitemOpen
  \bibfield  {author} {\bibinfo {author} {\bibfnamefont {R.}~\bibnamefont
  {Ozawa}}, \bibinfo {author} {\bibfnamefont {S.}~\bibnamefont {Hayami}},
  \bibinfo {author} {\bibfnamefont {K.}~\bibnamefont {Barros}}, \bibinfo
  {author} {\bibfnamefont {G.~W.}\ \bibnamefont {Chern}}, \bibinfo {author}
  {\bibfnamefont {Y.}~\bibnamefont {Motome}},\ and\ \bibinfo {author}
  {\bibfnamefont {C.~D.}\ \bibnamefont {Batista}},\ }\bibfield  {title}
  {\bibinfo {title} {{Vortex crystals with chiral stripes in itinerant
  magnets}},\ }\href {https://doi.org/10.7566/JPSJ.85.103703} {\bibfield
  {journal} {\bibinfo  {journal} {J. Phys. Soc. Jpn.}\ }\textbf {\bibinfo
  {volume} {85}},\ \bibinfo {pages} {103703} (\bibinfo {year}
  {2016})}\BibitemShut {NoStop}%
\bibitem [{\citenamefont {Hayami}\ \emph {et~al.}(2017)\citenamefont {Hayami},
  \citenamefont {Ozawa},\ and\ \citenamefont {Motome}}]{Hayami2017}%
  \BibitemOpen
  \bibfield  {author} {\bibinfo {author} {\bibfnamefont {S.}~\bibnamefont
  {Hayami}}, \bibinfo {author} {\bibfnamefont {R.}~\bibnamefont {Ozawa}},\ and\
  \bibinfo {author} {\bibfnamefont {Y.}~\bibnamefont {Motome}},\ }\bibfield
  {title} {\bibinfo {title} {{Effective bilinear-biquadratic model for
  noncoplanar ordering in itinerant magnets}},\ }\href
  {https://doi.org/10.1103/PhysRevB.95.224424} {\bibfield  {journal} {\bibinfo
  {journal} {Phys. Rev. B}\ }\textbf {\bibinfo {volume} {95}},\ \bibinfo
  {pages} {224424} (\bibinfo {year} {2017})}\BibitemShut {NoStop}%
\bibitem [{\citenamefont {Hayami}\ and\ \citenamefont
  {Motome}(2021{\natexlab{a}})}]{Hayami2021topological}%
  \BibitemOpen
  \bibfield  {author} {\bibinfo {author} {\bibfnamefont {S.}~\bibnamefont
  {Hayami}}\ and\ \bibinfo {author} {\bibfnamefont {Y.}~\bibnamefont
  {Motome}},\ }\bibfield  {title} {\bibinfo {title} {Topological spin crystals
  by itinerant frustration},\ }\href {https://doi.org/10.1088/1361-648X/ac1a30}
  {\bibfield  {journal} {\bibinfo  {journal} {Journal of Physics: Condensed
  Matter}\ }\textbf {\bibinfo {volume} {33}},\ \bibinfo {pages} {443001}
  (\bibinfo {year} {2021}{\natexlab{a}})}\BibitemShut {NoStop}%
\bibitem [{\citenamefont {Bray}(1994)}]{Bray1994}%
  \BibitemOpen
  \bibfield  {author} {\bibinfo {author} {\bibfnamefont {A.~J.}\ \bibnamefont
  {Bray}},\ }\bibfield  {title} {\bibinfo {title} {Theory of phase-ordering
  kinetics},\ }\href {https://doi.org/10.1080/00018739400101505} {\bibfield
  {journal} {\bibinfo  {journal} {Advances in Physics}\ }\textbf {\bibinfo
  {volume} {43}},\ \bibinfo {pages} {357} (\bibinfo {year} {1994})}\BibitemShut
  {NoStop}%
\bibitem [{\citenamefont {Puri}\ and\ \citenamefont
  {Wadhawan}(2009)}]{Puri2009}%
  \BibitemOpen
  \bibfield  {author} {\bibinfo {author} {\bibfnamefont {S.}~\bibnamefont
  {Puri}}\ and\ \bibinfo {author} {\bibfnamefont {V.}~\bibnamefont
  {Wadhawan}},\ }\href@noop {} {\emph {\bibinfo {title} {Kinetics of phase
  transitions}}}\ (\bibinfo  {publisher} {CRC press},\ \bibinfo {year}
  {2009})\BibitemShut {NoStop}%
\bibitem [{\citenamefont {Ruderman}\ and\ \citenamefont
  {Kittel}(1954)}]{Ruderman1954}%
  \BibitemOpen
  \bibfield  {author} {\bibinfo {author} {\bibfnamefont {M.~A.}\ \bibnamefont
  {Ruderman}}\ and\ \bibinfo {author} {\bibfnamefont {C.}~\bibnamefont
  {Kittel}},\ }\bibfield  {title} {\bibinfo {title} {{Indirect Exchange
  Coupling of Nuclear Magnetic Moments by Conduction Electrons}},\ }\href
  {https://doi.org/10.1103/PhysRev.96.99} {\bibfield  {journal} {\bibinfo
  {journal} {Phys. Rev.}\ }\textbf {\bibinfo {volume} {96}},\ \bibinfo {pages}
  {99} (\bibinfo {year} {1954})}\BibitemShut {NoStop}%
\bibitem [{\citenamefont {Kasuya}(1956)}]{Kasuya1956}%
  \BibitemOpen
  \bibfield  {author} {\bibinfo {author} {\bibfnamefont {T.}~\bibnamefont
  {Kasuya}},\ }\bibfield  {title} {\bibinfo {title} {{A Theory of Metallic
  Ferro- and Antiferromagnetism on Zener's Model}},\ }\href
  {https://doi.org/10.1143/PTP.16.45} {\bibfield  {journal} {\bibinfo
  {journal} {Prog. Theor. Phys.}\ }\textbf {\bibinfo {volume} {16}},\ \bibinfo
  {pages} {45} (\bibinfo {year} {1956})}\BibitemShut {NoStop}%
\bibitem [{\citenamefont {Yosida}(1957)}]{Yosida1957}%
  \BibitemOpen
  \bibfield  {author} {\bibinfo {author} {\bibfnamefont {K.}~\bibnamefont
  {Yosida}},\ }\bibfield  {title} {\bibinfo {title} {Magnetic properties of
  cu-mn alloys},\ }\href {https://doi.org/10.1103/PhysRev.106.893} {\bibfield
  {journal} {\bibinfo  {journal} {Phys. Rev.}\ }\textbf {\bibinfo {volume}
  {106}},\ \bibinfo {pages} {893} (\bibinfo {year} {1957})}\BibitemShut
  {NoStop}%
\bibitem [{\citenamefont {Hayami}\ and\ \citenamefont
  {Motome}(2018)}]{Hayami2018}%
  \BibitemOpen
  \bibfield  {author} {\bibinfo {author} {\bibfnamefont {S.}~\bibnamefont
  {Hayami}}\ and\ \bibinfo {author} {\bibfnamefont {Y.}~\bibnamefont
  {Motome}},\ }\bibfield  {title} {\bibinfo {title} {{N\'eel- and Bloch-Type
  Magnetic Vortices in Rashba Metals}},\ }\href
  {https://doi.org/10.1103/PhysRevLett.121.137202} {\bibfield  {journal}
  {\bibinfo  {journal} {Phys. Rev. Lett.}\ }\textbf {\bibinfo {volume} {121}},\
  \bibinfo {pages} {137202} (\bibinfo {year} {2018})}\BibitemShut {NoStop}%
\bibitem [{1_s()}]{1_supp_Heff}%
  \BibitemOpen
  \href@noop {} {}\bibinfo {note} {A detailed discussion of the effective
  Hamiltonian for the square-lattice s-d model with Rashba SOC is presented in
  the supplemental material.}\BibitemShut {Stop}%
\bibitem [{\citenamefont {Okumura}\ \emph {et~al.}(2020)\citenamefont
  {Okumura}, \citenamefont {Hayami}, \citenamefont {Kato},\ and\ \citenamefont
  {Motome}}]{Okumura2020}%
  \BibitemOpen
  \bibfield  {author} {\bibinfo {author} {\bibfnamefont {S.}~\bibnamefont
  {Okumura}}, \bibinfo {author} {\bibfnamefont {S.}~\bibnamefont {Hayami}},
  \bibinfo {author} {\bibfnamefont {Y.}~\bibnamefont {Kato}},\ and\ \bibinfo
  {author} {\bibfnamefont {Y.}~\bibnamefont {Motome}},\ }\bibfield  {title}
  {\bibinfo {title} {{Magnetic hedgehog lattices in noncentrosymmetric
  metals}},\ }\href {https://doi.org/10.1103/PhysRevB.101.144416} {\bibfield
  {journal} {\bibinfo  {journal} {Phys. Rev. B}\ }\textbf {\bibinfo {volume}
  {101}},\ \bibinfo {pages} {144416} (\bibinfo {year} {2020})}\BibitemShut
  {NoStop}%
\bibitem [{\citenamefont {Yambe}\ and\ \citenamefont
  {Hayami}(2021)}]{Yambe2021skyrmion}%
  \BibitemOpen
  \bibfield  {author} {\bibinfo {author} {\bibfnamefont {R.}~\bibnamefont
  {Yambe}}\ and\ \bibinfo {author} {\bibfnamefont {S.}~\bibnamefont {Hayami}},\
  }\bibfield  {title} {\bibinfo {title} {Skyrmion crystals in centrosymmetric
  itinerant magnets without horizontal mirror plane},\ }\href@noop {}
  {\bibfield  {journal} {\bibinfo  {journal} {Sci. Rep.}\ }\textbf {\bibinfo
  {volume} {11}},\ \bibinfo {pages} {11184} (\bibinfo {year}
  {2021})}\BibitemShut {NoStop}%
\bibitem [{\citenamefont {Shimizu}\ \emph {et~al.}(2021)\citenamefont
  {Shimizu}, \citenamefont {Okumura}, \citenamefont {Kato},\ and\ \citenamefont
  {Motome}}]{Shimizu2021anisotropy}%
  \BibitemOpen
  \bibfield  {author} {\bibinfo {author} {\bibfnamefont {K.}~\bibnamefont
  {Shimizu}}, \bibinfo {author} {\bibfnamefont {S.}~\bibnamefont {Okumura}},
  \bibinfo {author} {\bibfnamefont {Y.}~\bibnamefont {Kato}},\ and\ \bibinfo
  {author} {\bibfnamefont {Y.}~\bibnamefont {Motome}},\ }\bibfield  {title}
  {\bibinfo {title} {{Phase transitions between helices, vortices, and
  hedgehogs driven by spatial anisotropy in chiral magnets}},\ }\href
  {https://doi.org/10.1103/PhysRevB.103.054427} {\bibfield  {journal} {\bibinfo
   {journal} {Phys. Rev. B}\ }\textbf {\bibinfo {volume} {103}},\ \bibinfo
  {pages} {054427} (\bibinfo {year} {2021})}\BibitemShut {NoStop}%
\bibitem [{\citenamefont {Kato}\ \emph {et~al.}(2021)\citenamefont {Kato},
  \citenamefont {Hayami},\ and\ \citenamefont {Motome}}]{Kato2021}%
  \BibitemOpen
  \bibfield  {author} {\bibinfo {author} {\bibfnamefont {Y.}~\bibnamefont
  {Kato}}, \bibinfo {author} {\bibfnamefont {S.}~\bibnamefont {Hayami}},\ and\
  \bibinfo {author} {\bibfnamefont {Y.}~\bibnamefont {Motome}},\ }\bibfield
  {title} {\bibinfo {title} {{Spin excitation spectra in helimagnetic states:
  Proper-screw, cycloid, vortex-crystal, and hedgehog lattices}},\ }\href
  {https://doi.org/10.1103/PhysRevB.104.224405} {\bibfield  {journal} {\bibinfo
   {journal} {Phys. Rev. B}\ }\textbf {\bibinfo {volume} {104}},\ \bibinfo
  {pages} {224405} (\bibinfo {year} {2021})}\BibitemShut {NoStop}%
\bibitem [{\citenamefont {Yurke}\ \emph {et~al.}(1993)\citenamefont {Yurke},
  \citenamefont {Pargellis}, \citenamefont {Kovacs},\ and\ \citenamefont
  {Huse}}]{Yurke1993}%
  \BibitemOpen
  \bibfield  {author} {\bibinfo {author} {\bibfnamefont {B.}~\bibnamefont
  {Yurke}}, \bibinfo {author} {\bibfnamefont {A.~N.}\ \bibnamefont
  {Pargellis}}, \bibinfo {author} {\bibfnamefont {T.}~\bibnamefont {Kovacs}},\
  and\ \bibinfo {author} {\bibfnamefont {D.~A.}\ \bibnamefont {Huse}},\
  }\bibfield  {title} {\bibinfo {title} {Coarsening dynamics of the xy model},\
  }\href {https://doi.org/10.1103/PhysRevE.47.1525} {\bibfield  {journal}
  {\bibinfo  {journal} {Phys. Rev. E}\ }\textbf {\bibinfo {volume} {47}},\
  \bibinfo {pages} {1525} (\bibinfo {year} {1993})}\BibitemShut {NoStop}%
\bibitem [{\citenamefont {Bray}\ \emph {et~al.}(2000)\citenamefont {Bray},
  \citenamefont {Briant},\ and\ \citenamefont {Jervis}}]{Bray2000}%
  \BibitemOpen
  \bibfield  {author} {\bibinfo {author} {\bibfnamefont {A.~J.}\ \bibnamefont
  {Bray}}, \bibinfo {author} {\bibfnamefont {A.~J.}\ \bibnamefont {Briant}},\
  and\ \bibinfo {author} {\bibfnamefont {D.~K.}\ \bibnamefont {Jervis}},\
  }\bibfield  {title} {\bibinfo {title} {Breakdown of scaling in the
  nonequilibrium critical dynamics of the two-dimensional $\mathit{XY}$
  model},\ }\href {https://doi.org/10.1103/PhysRevLett.84.1503} {\bibfield
  {journal} {\bibinfo  {journal} {Phys. Rev. Lett.}\ }\textbf {\bibinfo
  {volume} {84}},\ \bibinfo {pages} {1503} (\bibinfo {year}
  {2000})}\BibitemShut {NoStop}%
\bibitem [{2_s()}]{2_supp_dislocation}%
  \BibitemOpen
  \href@noop {} {}\bibinfo {note} {A detailed discussion of the way to count
  the dislocations and the relationship between dislocations in the present
  system and vortices in the XY model are presented in the supplemental
  material.}\BibitemShut {Stop}%
\bibitem [{\citenamefont {Algara-Siller}\ \emph {et~al.}(2015)\citenamefont
  {Algara-Siller}, \citenamefont {Lehtinen}, \citenamefont {Wang},
  \citenamefont {Nair}, \citenamefont {Kaiser}, \citenamefont {Wu},
  \citenamefont {Geim},\ and\ \citenamefont {Grigorieva}}]{Siller2015}%
  \BibitemOpen
  \bibfield  {author} {\bibinfo {author} {\bibfnamefont {G.}~\bibnamefont
  {Algara-Siller}}, \bibinfo {author} {\bibfnamefont {O.}~\bibnamefont
  {Lehtinen}}, \bibinfo {author} {\bibfnamefont {F.~C.}\ \bibnamefont {Wang}},
  \bibinfo {author} {\bibfnamefont {R.~R.}\ \bibnamefont {Nair}}, \bibinfo
  {author} {\bibfnamefont {U.}~\bibnamefont {Kaiser}}, \bibinfo {author}
  {\bibfnamefont {H.~A.}\ \bibnamefont {Wu}}, \bibinfo {author} {\bibfnamefont
  {A.~K.}\ \bibnamefont {Geim}},\ and\ \bibinfo {author} {\bibfnamefont
  {I.~V.}\ \bibnamefont {Grigorieva}},\ }\bibfield  {title} {\bibinfo {title}
  {Square ice in graphene nanocapillaries},\ }\href
  {https://doi.org/10.1038/nature14295} {\bibfield  {journal} {\bibinfo
  {journal} {Nature}\ }\textbf {\bibinfo {volume} {519}},\ \bibinfo {pages}
  {443} (\bibinfo {year} {2015})}\BibitemShut {NoStop}%
\bibitem [{\citenamefont {Khanh}\ \emph {et~al.}(2020)\citenamefont {Khanh},
  \citenamefont {Nakajima}, \citenamefont {Yu}, \citenamefont {Gao},
  \citenamefont {Shibata}, \citenamefont {Hirschberger}, \citenamefont
  {Yamasaki}, \citenamefont {Sagayama}, \citenamefont {Nakao}, \citenamefont
  {Peng}, \citenamefont {Nakajima}, \citenamefont {Takagi}, \citenamefont
  {Arima}, \citenamefont {Tokura},\ and\ \citenamefont {Seki}}]{Khanh2020}%
  \BibitemOpen
  \bibfield  {author} {\bibinfo {author} {\bibfnamefont {N.~D.}\ \bibnamefont
  {Khanh}}, \bibinfo {author} {\bibfnamefont {T.}~\bibnamefont {Nakajima}},
  \bibinfo {author} {\bibfnamefont {X.}~\bibnamefont {Yu}}, \bibinfo {author}
  {\bibfnamefont {S.}~\bibnamefont {Gao}}, \bibinfo {author} {\bibfnamefont
  {K.}~\bibnamefont {Shibata}}, \bibinfo {author} {\bibfnamefont
  {M.}~\bibnamefont {Hirschberger}}, \bibinfo {author} {\bibfnamefont
  {Y.}~\bibnamefont {Yamasaki}}, \bibinfo {author} {\bibfnamefont
  {H.}~\bibnamefont {Sagayama}}, \bibinfo {author} {\bibfnamefont
  {H.}~\bibnamefont {Nakao}}, \bibinfo {author} {\bibfnamefont
  {L.}~\bibnamefont {Peng}}, \bibinfo {author} {\bibfnamefont {K.}~\bibnamefont
  {Nakajima}}, \bibinfo {author} {\bibfnamefont {R.}~\bibnamefont {Takagi}},
  \bibinfo {author} {\bibfnamefont {T.}~\bibnamefont {Arima}}, \bibinfo
  {author} {\bibfnamefont {Y.}~\bibnamefont {Tokura}},\ and\ \bibinfo {author}
  {\bibfnamefont {S.}~\bibnamefont {Seki}},\ }\bibfield  {title} {\bibinfo
  {title} {{Nanometric square skyrmion lattice in a centrosymmetric tetragonal
  magnet}},\ }\href {https://doi.org/10.1038/s41565-020-0684-7} {\bibfield
  {journal} {\bibinfo  {journal} {Nat. Nanotechnol.}\ }\textbf {\bibinfo
  {volume} {15}},\ \bibinfo {pages} {444} (\bibinfo {year} {2020})}\BibitemShut
  {NoStop}%
\bibitem [{\citenamefont {Takagi}\ \emph {et~al.}(2022)\citenamefont {Takagi},
  \citenamefont {Matsuyama}, \citenamefont {Ukleev}, \citenamefont {Yu},
  \citenamefont {White}, \citenamefont {Francoual}, \citenamefont {Mardegan},
  \citenamefont {Hayami}, \citenamefont {Saito}, \citenamefont {Kaneko} \emph
  {et~al.}}]{Takagi2022}%
  \BibitemOpen
  \bibfield  {author} {\bibinfo {author} {\bibfnamefont {R.}~\bibnamefont
  {Takagi}}, \bibinfo {author} {\bibfnamefont {N.}~\bibnamefont {Matsuyama}},
  \bibinfo {author} {\bibfnamefont {V.}~\bibnamefont {Ukleev}}, \bibinfo
  {author} {\bibfnamefont {L.}~\bibnamefont {Yu}}, \bibinfo {author}
  {\bibfnamefont {J.~S.}\ \bibnamefont {White}}, \bibinfo {author}
  {\bibfnamefont {S.}~\bibnamefont {Francoual}}, \bibinfo {author}
  {\bibfnamefont {J.~R.}\ \bibnamefont {Mardegan}}, \bibinfo {author}
  {\bibfnamefont {S.}~\bibnamefont {Hayami}}, \bibinfo {author} {\bibfnamefont
  {H.}~\bibnamefont {Saito}}, \bibinfo {author} {\bibfnamefont
  {K.}~\bibnamefont {Kaneko}}, \emph {et~al.},\ }\bibfield  {title} {\bibinfo
  {title} {Square and rhombic lattices of magnetic skyrmions in a
  centrosymmetric binary compound},\ }\href@noop {} {\bibfield  {journal}
  {\bibinfo  {journal} {Nat. Commun.}\ }\textbf {\bibinfo {volume} {13}},\
  \bibinfo {pages} {1472} (\bibinfo {year} {2022})}\BibitemShut {NoStop}%
\bibitem [{\citenamefont {Hayami}\ and\ \citenamefont
  {Motome}(2021{\natexlab{b}})}]{Hayami2021square}%
  \BibitemOpen
  \bibfield  {author} {\bibinfo {author} {\bibfnamefont {S.}~\bibnamefont
  {Hayami}}\ and\ \bibinfo {author} {\bibfnamefont {Y.}~\bibnamefont
  {Motome}},\ }\bibfield  {title} {\bibinfo {title} {Square skyrmion crystal in
  centrosymmetric itinerant magnets},\ }\href
  {https://doi.org/10.1103/PhysRevB.103.024439} {\bibfield  {journal} {\bibinfo
   {journal} {Phys. Rev. B}\ }\textbf {\bibinfo {volume} {103}},\ \bibinfo
  {pages} {024439} (\bibinfo {year} {2021}{\natexlab{b}})}\BibitemShut
  {NoStop}%
\end{thebibliography}%

\end{document}